\documentclass[pra,aps,twocolumn]{revtex4-1}
\usepackage{amsmath}
\usepackage{amssymb}
\usepackage{graphicx}
\usepackage{bbold}
\usepackage{csquotes}

\newcommand{\beqa}{\begin{eqnarray}}
\newcommand{\eeqa}{\end{eqnarray}}

\begin{document}

\title{Lambert function methods for laser dynamics with time-delayed feedback}

\author{Yogesh N. Joglekar}
\author{Andrew Wilkey}
\author{Gautam Vemuri}

\affiliation{Department of Physics, Indiana University Purdue University Indianapolis (IUPUI), Indianapolis 46202 Indiana, USA}

\begin{abstract}
Time-delayed differential equations arise frequently in the study of nonlinear dynamics of lasers with optical feedback. Traditionally, one has resorted to numerical methods because the analytical solution of such equations are intractable.  In this manuscript, we show that under some conditions, the rate equations model that is used to model semiconductor lasers with feedback can be analytically solved by using the Lambert W function.  In particular, we discuss the conditions under which the coupled rate equations for the intra-cavity electric field and excess carrier inversion can be reduced to a single equation for the field, and how this single rate equation can be cast in a form that is amenable to the use of the Lambert W function. We conclude the manuscript with a similar discussion for two lasers coupled via time-delayed feedbacks. 
\end{abstract}
\maketitle

%----------------- --------------------------------------------------------------------%

\section{Introduction}

Time-delayed differential equations arise naturally in a wide variety of physical phenomena where one ore more system parameters are fed back into the system after a certain amount of time.  Such time-delayed feedbacks are seen in population behaviors in biology and ecology~\cite{smith,blythe,kuang}, chemical reactions~\cite{marc}, interactions between time-delayed non-Markovian laser fields and resonant media~\cite{anderson} and a host of nonlinear dynamical systems~\cite{lenstra}. One of the more prominent examples of a physical system with time-delayed feedback is a laser wherein the light emitted by a laser is injected back into the laser by reflection from a distant mirror outside the laser cavity~\cite{kane}. The mathematical model for a time-delayed feedback system often reduces to a first order differential equation with a time-delayed term, and the analytical solution of such differential equations can be difficult because one has to deal with an infinite-dimensional equation.  In this article, we demonstrate that the Lambert W function~\cite{corless} can be invoked in some situations to obtain analytical solutions to time-delayed equations of physical interest, and explore some of the consequences of using this method.

For the sake of concreteness, we focus on the problem of a semiconductor laser that is subject to time-delayed feedback of light into the laser~\cite{van}. Lasers with time-delayed feedback are a paradigm for the study of time-delayed systems in part because the delay can be easily controlled, which allows one to study the behavior of the system for delays that are shorter than the intrinsic time-scales of the laser as well as for delays that are longer than the natural time scales of the isolated laser system. Such lasers are of fundamental interest due to the variety of nonlinear dynamical behaviors that arise as a function of the time-delay and the strength of the feedback~\cite{mork}. In particular, there are combinations of delay and feedback strengths that produce single-tone oscillations in the optical frequency of the laser, period doubling routes to chaos, and coherence collapse and line-narrowing. Each of these dynamical responses have been studied for a variety of applications such as the development of stable, all-optical microwave frequency oscillators, chaotic synchronization for all-optical encryption, and stable, narrow line-width lasers~\cite{soriano}. Another system that has been of immense interest to the semiconductor laser community is the coupling of two lasers by mutual injection of light from each laser into the other~\cite{mulet}. These systems have a natural time-delay built into them due to the finite amount of time it takes for the light from one laser to reach the other laser due to the physical separation between the lasers.

%----------------- --------------------------------------------------------------------%

\section{Lang Kobayashi equations}
\label{sec:lke}

%--------------------------------------------------------------------------------------%
\begin{figure*}[htb]
\includegraphics[width=\columnwidth]{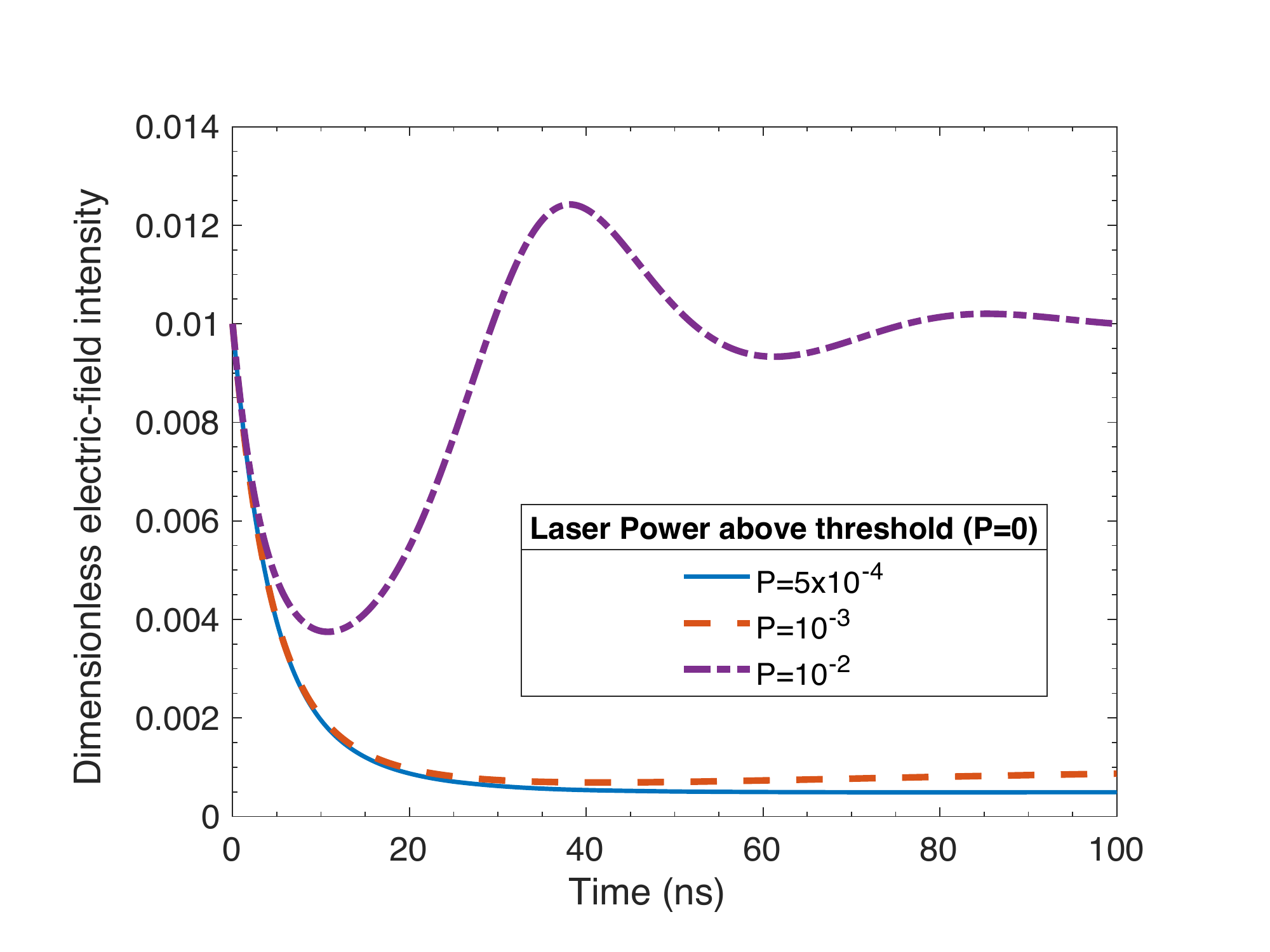}
\includegraphics[width=\columnwidth]{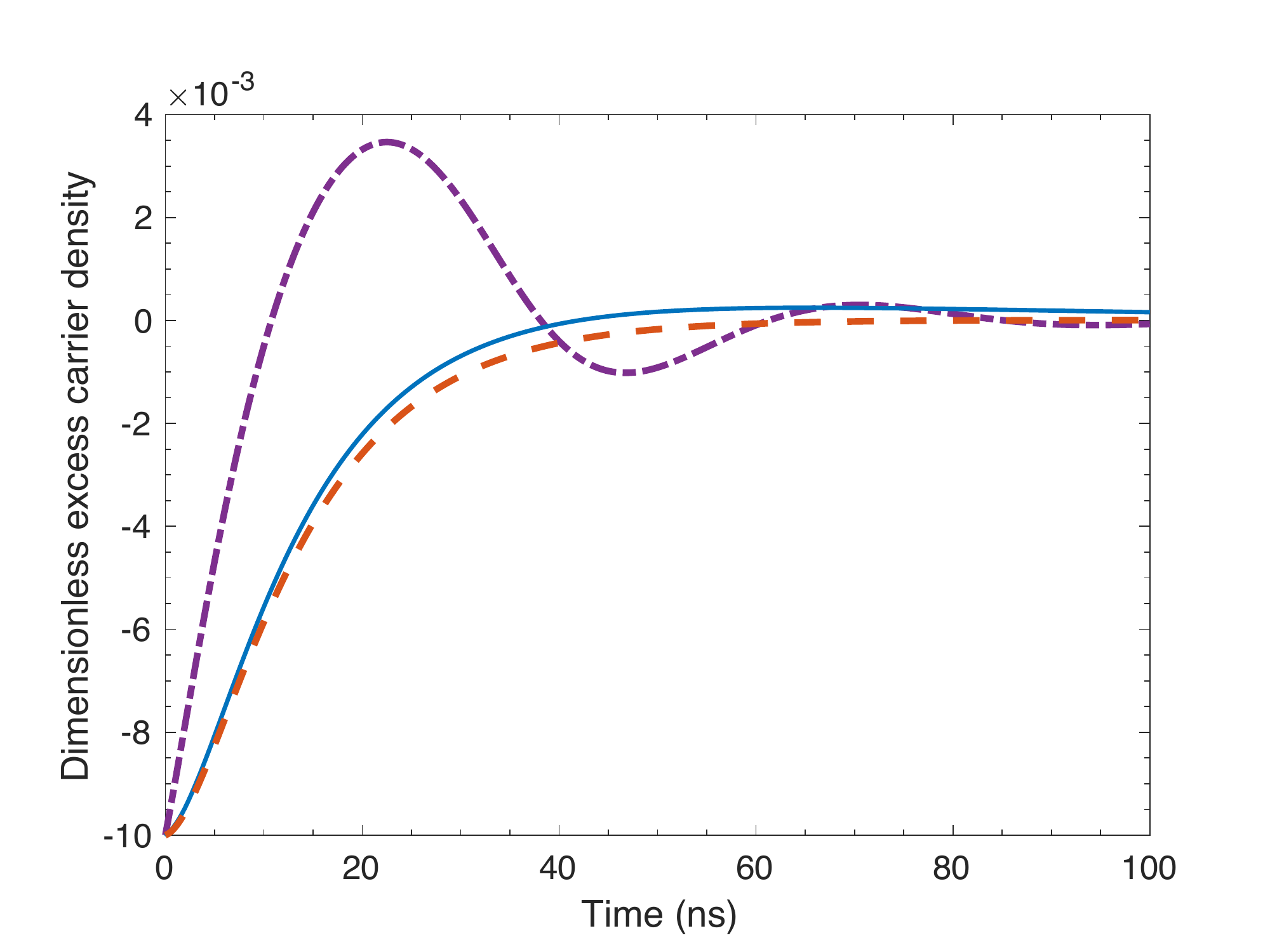}
\caption{Full temporal dynamics of Electric field intensity $I(t)=|E(t)|^2$ and excess carrier inversion $N(t)$ in a typical semiconductor laser with line-width enhancement factor $\alpha=3$ and $T=100$, as a function of the pump power $P$ above the threshold. Note that both dynamics have the same time scale, and change from overdamped to underdamped in their approach to equilibrium as the pump power increases.}
\label{fig:fullLK}
\end{figure*}
%--------------------------------------------------------------------------------------%

Semiconductor lasers are usually modeled by the Lang-Kobayashi equations~\cite{lang} which are known to describe the experimentally observed behavior of these lasers very well.  For a single-mode laser,  these equations describe the coupled time evolution of the electric field and the excess carrier inversion inside a laser cavity. In the slowly-varying envelope approximation, these equations in the non-dimensional form are 

\begin{eqnarray}
\frac{dE}{dt} & =& (1+i\alpha){\zeta}N(t)E(t)+{\kappa}E(t-{\tau}),\label{eq:lk1}\\
T\frac{dN}{dt} & = & P-N(t)-(1+2N(t))|E(t)|^{2}.\label{eq:lk2}
\end{eqnarray}

Here $E(t)$ is the complex, time-dependent, intra-cavity electric field, $\alpha$ is the line-width enhancement factor for the gain medium, $N(t)$ is the time-dependent excess carrier inversion (above the carrier inversion at the lasing threshold), $\zeta$ is the differential-gain coefficient, $\kappa$ is the feedback coupling strength, $\tau$ is the time-delay, $T$ is the ratio of excess carrier-inversion lifetime to the cavity photon lifetime, and $P$ is the external pumping to the laser.  Note that the rate equation for the macroscopic polarization within the gain medium does not enter this model because it decays very rapidly, relative to the time scale at which $E(t)$ and $N(t)$ evolve in semiconductor lasers, and hence can be adiabatically eliminated. Figure~\ref{fig:fullLK} shows the results for the envelope field intensity $|E(t)|^2$ and the excess carrier inversion $N(t)$ as a function of the pump power above the lasing threshold. When the pump power is small, both intensity and excess inversion approach their equilibrium values rapidly in an overdamped manner. As the pump power increases, the approach to the equilibrium values changes to an underdamped manner. In dissipative systems, this transition from overdamped to underdamped approach can be mapped onto a parity-time ($\mathcal{PT}$) symmetry breaking transition~\cite{leluo} where the overdamped region is associated with $\mathcal{PT}$-broken phase and the underdamped region is associated with the $\mathcal{PT}$-symmetric phase. For realistic parameters of standard semiconductor lasers, used in Fig.~\ref{fig:fullLK}, we note that the time scales for variation in the electric field and the excess inversion are the same. Thus, it is not possible to eliminate the excess carrier density dynamics in the current set up, and the Lambert function formalism cannot be directly applied to above set of equations~(\ref{eq:lk1})-(\ref{eq:lk2}).

The Lambert W-function is defined by solutions of the equation $we^w=z$. For a general complex number $z$, this equation has countably infinite number of solutions denoted by $W_k(z)$ for integers $k$, out of which, by convention, only branches $k=0$ and $k=-1$ are real-valued for any $z$~\cite{corless}. Suppose we are able to reduce the Lang-Kobayashi equations to a single equation of the form 
\begin{equation}
\label{eq:lambert0}
\frac{dx}{dt}=ax+bx(t-{\tau})
\end{equation}
where $x$ is complex and $a, b$ are constants that could be real or complex. We note that such equations commonly arise in time-delayed population dynamics models in biology and ecology~\cite{smith,blythe,kuang}, but in those cases, the coefficients and the solution of the equation are both constrained to be purely real. Since Eq.(\ref{eq:lambert0} is a {\it linear equation}, its general solution will be given by the linear superposition of exponential-in-time terms. Assuming $x(t)\sim e^{\lambda t}$ leads to a transcendental characteristic equation 
\begin{equation}
\label{eq:lambert1}
(\lambda-a)\tau e^{(\lambda-a)\tau}=b\tau e^{a\tau}.
\end{equation}
Equation~\ref{eq:lambert1} shows that the eigenvalues $\lambda$ can be expressed in terms of the Lambert $W$-function. Specifically, when if $b{\tau}e^{{\tau}a} >0$, the solution for is given by $\lambda=a+W_{0}(b{\tau}e^{a\tau})/\tau$. If $b{\tau}e^{a\tau} <0$, there are two possible solutions when $-1/e<b{\tau}e^{a\tau}<0$, one smaller than 1 and the other larger than 1.  Thus, in general, the eigenvalues $\lambda$ that characterize the exponential-in-time behavior of $x(t)$ are given by 
\begin{equation}
\label{eq:lambert2}
\lambda_k=a+\frac{1}{\tau}W_{k}(b{\tau}e^{a\tau})
\end{equation}
where for $b{\tau}e^{a\tau} <-1/e$ or complex, the general solution is obtained by an analytical continuation of the function to the complex plane~\cite{corless}. 

\section{Domain of validity of Lambert formalism}

\begin{figure}
\includegraphics[width=\columnwidth]{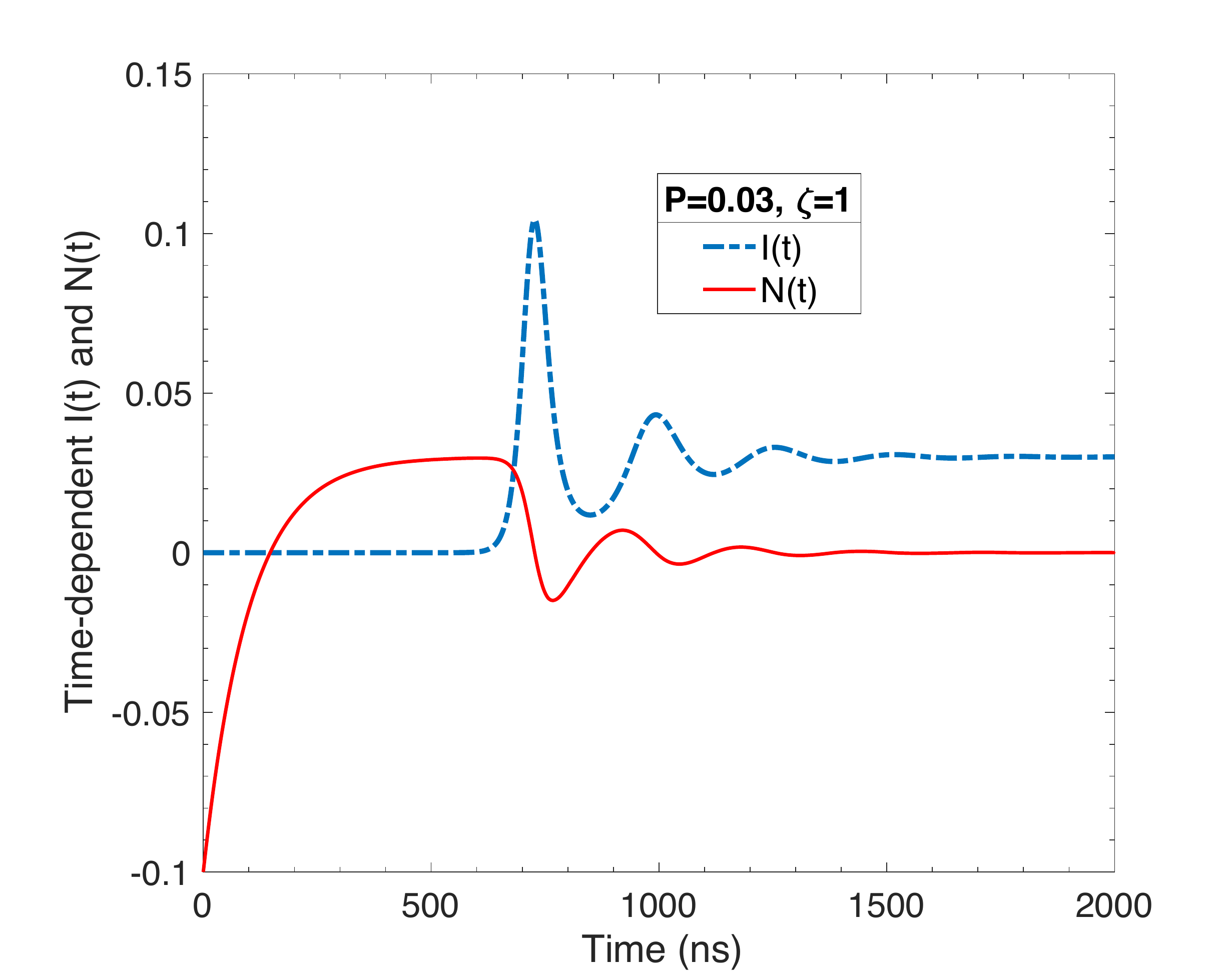}
\caption{Time evolution of the dimensionless electric field intensity $I(t)=|E(t)|^2$ and the excess carrier inversion $N(t)$ in typical, solitary semiconductor laser shows that the electric field rises exponentially in a very short window before nonlinearities in the gain medium become effective and saturate the intensity to a steady state value.}
\label{fig:ne}
\end{figure}

It is generally the case that the modeling of semiconductor lasers requires the coupled rate equations for the intracavity electric field, $E(t)$ and the excess carrier inversion, $N(t)$.  However, the use of the Lambert equation formalism, as discussed above, requires that the two coupled rate equations be reduced to a single equation with time-delay.  In this case, it means that the model must be reduced to a rate equation for the electric field only.  This is, strictly speaking, not possible for semiconductor lasers because, as seen in Fig.~\ref{fig:fullLK}, the characteristic time scales on which the intracavity field and the excess carrier inversion evolve are of the same order of magnitude.  

However, with advancements in technology, it may well be possible to fabricate lasers in which the excess   carrier inversion evolves much faster than the electric field, in which case the rate equation for the excess carrier inversion can be eliminated.  This entails setting the time evolution of the inversion to zero, i.e. $dN(t)/dt = 0$, solving for the steady state value of $N(t)$, and substituting this expression for $N(t)$ in the rate equation for the electric field, Eq.(\ref{eq:lk1}).  The model is thereby reduced to a single, time-delayed rate equation that is amenable to the use of the Lambert function.  Another possibility is to fabricate a laser in which the carrier inversion relaxation time is much larger than the electric field decay time.  In this case, the inversion does not evolve during the time that the electric field reaches a steady state, and one can, once again, focus on the single, time-delayed rate equation for the electric field.

In addition to the elimination of the inversion equation, the Lambert formalism is strictly applicable to the case of a semiconductor laser with time-delayed feedback only when the nonlinearities that arise from the cubic term in the electric field are also ignored. Physically, this means that the formalism is valid when two conditions are simultaneously satisfied - (i) the excess carrier inversion $N(t)$ has reached its steady state value, and (ii) the laser has reaches threshold and the intracavity electric field $E(t)$ is starting to grow exponentially, but the intensity saturation that typically sets in due to the nonlinearities in the gain is yet to occur.  For typical semiconductor laser parameters (with no time-delayed feedback), this time window between the inversion settling to a steady state value and the intensity getting saturated is negligibly small. Figure~\ref{fig:ne} shows the typical time evolution of a solitary semiconductor laser ($\alpha=3$, pumping $P=0.03$, and differential gain coefficient $\zeta=1$) with initial excess carrier inversion, i.e. $N(t=0)=-0.1$, and we see that the exponential electric field rise occurs in a very short time interval. This essentially means that the predictions of the analytic solutions, Eq.(\ref{eq:lambert2}), obtained via the Lambert function approach will be visible in a very tiny time window. 

%--------------------------------------------------------------------------------------%
\begin{figure*}[thb]
\includegraphics[width=\columnwidth]{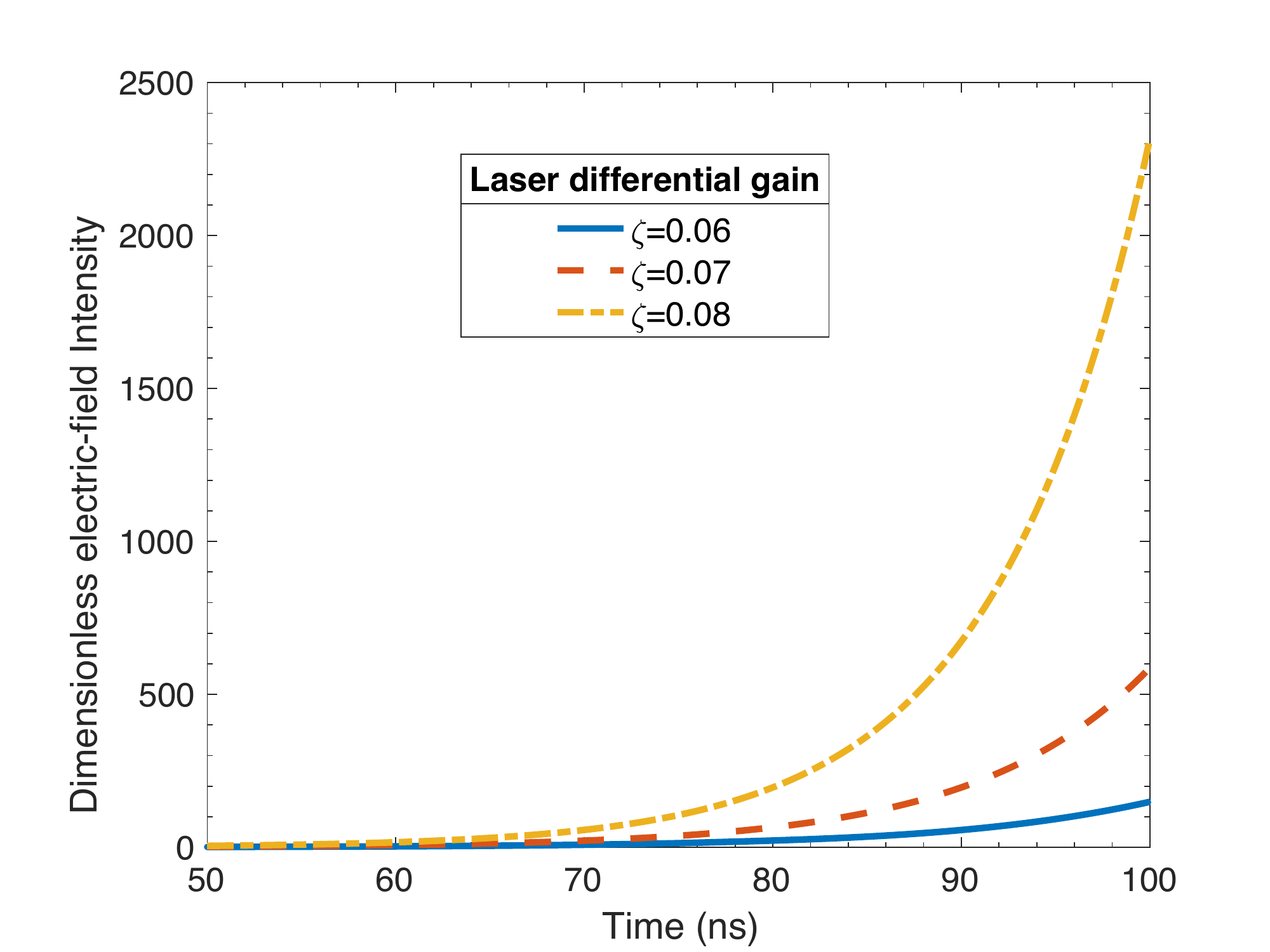}
\includegraphics[width=\columnwidth]{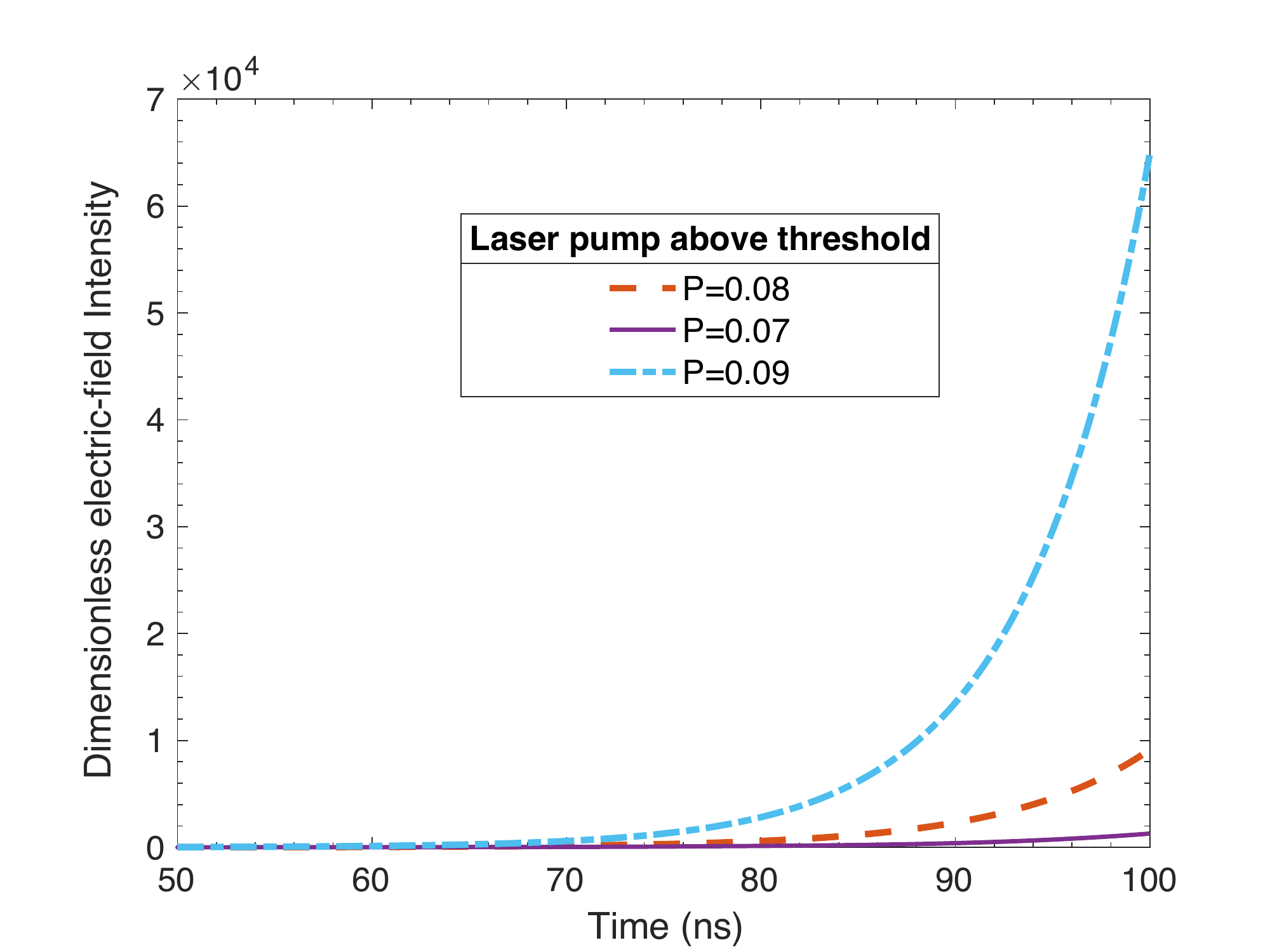}
\caption{Rise of the intracavity electric field $E(t)$ when the carrier inversion is adiabatically eliminated. The left-hand panel shows that for a fixed $\zeta=0.1$, as the pump current is reduced, the exponential growth of intensity slows down. The right-hand panel shows that for a fixed pumping $P=0.08$, as $\zeta$ is reduced, the time window during which the gain-nonlinearity can be ignored is widened. In both cases, the calculations are carried out using typical semiconductor laser parameters, $\alpha=3$. Note that the time axis only contains time window after the carrier inversion has reached steady-state value.}
\label{fig:lam}
\end{figure*}
%--------------------------------------------------------------------------------------%

What is desired however is a wider time window between the inversion reaching a steady state and the intensity of the laser not yet being saturated, i.e. the intensity is still in the exponential amplification regime.  Since the intensity at very short times grows exponentially with the product of the differential gain coefficient $\zeta$ and $N(t)$, and because the steady-state value of $N(t)$ is a function of the pumping $P$, one can manipulate these two quantities to slow down the exponential growth of laser intensity and thereby enhance the time window in which both conditions are simultaneously met.  

Figure~\ref{fig:lam} shows the time-dependent laser intensities $I(t)=|E(t)|^2$ when the inversion is adiabatically eliminated and only the electric field equation is solved, as a function of differential gain $\zeta$ and pump power $P$. We note that the time axis only contains range after which the carrier inversion has settled down. It is evident from the left-hand panel that for a fixed $\zeta$, as the pump power is reduced from $P=0.09$ to $P=0.07$, the exponential growth of intensity is slowed, thereby providing a wider window of time in which to make any desired measurements of the laser intensity dynamics.  The right-hand panel shows that for a fixed pumping, as $\zeta$ is reduced, the intensity growth is slowed.  While the external pump is an easily varied parameter in experiments, the differential gain coefficient $\zeta$ is a material parameter and hence cannot be tuned in a given laser. On the other hand, reducing the pump power close to zero means the laser is operating very close to the threshold, and that leads to enhanced quantum fluctuations whose effects are not included in the present analysis. Thus, our results provide some guidance on the material parameters that are necessary for a laser to meet such that the Lambert formalism will be applicable to analytically study the dynamics of the laser.

Under these conditions, the rate equation for the electric field can be written as
\begin{equation}
\label{eq:eonly}
\frac{dE}{dt} = (1+i\alpha){\zeta}N_0E(t)+{\kappa}E(t-{\tau})
\end{equation}
where $N_0\sim P$ is the steady-state value of the carrier inversion. This equation is identical to Eq.(\ref{eq:lambert0}) with a manifestly complex $a=(1+i\alpha)\zeta N_0$ and a purely real, positive $b=\kappa$. Thus, within the appropriate time-window, the electric field exponents are determined by the properties of Lambert $W$-function. 

\section{Bidirectionally coupled lasers}

Another interesting situation in which the Lambert formalism can be invoked is when two identical semiconductor lasers, at optical frequencies $\omega_{1}$ and $\omega_{2}$ are mutually coupled to each other.  Such systems have been extensively studied in the context of their nonlinear dynamics~\cite{mork}. The four rate equations for such a system, two for the intracavity electric fields and two for the corresponding excess carrier inversions, are given by a modified form of the Lang-Kobayashi equations wherein the bidirectional coupling is accounted for, i.e., 
\begin{eqnarray}
\frac{dE_{1}(t)}{dt} & = & (1+i\alpha){\zeta}N_{1}(t)E_{1}(t)+i{\Delta}{\omega}E_1(t)\nonumber\\
&& +{\kappa}e^{-i{\Theta}{\tau}}E_{2}(t-\tau),\\
\frac{dE_{2}(t)}{dt} & = & (1+i\alpha){\zeta}N_{2}(t)E_{2}(t)-i{\Delta}{\omega}E_2(t)\nonumber\\
&& +{\kappa}e^{-i{\Theta}{\tau}}E_{1}(t-\tau),\\
T\frac{dN_{1}(t)}{dt} & = & J_{1}-N_{1}(t)\nonumber\\
&& -(1+2N_{1}(t))|E_{1}(t)|^{2},\\
T\frac{dN_{2}(t)}{dt} & = & J_{2}-N_{2}(t)\nonumber\\
&& -(1+2N_{2}(t))|E_{2}(t)|^{2}.
\end{eqnarray}
Here the subscripts 1,2 denote laser index,and these equations are written in a frame rotating at a frequency that is the average of the two laser frequencies, $\Theta =({\omega_{1}}+{\omega_{2}})/2$, so that each laser is detuned by an equal amount ${\pm}{\Delta}{\omega}=\pm(\omega_{1}-\omega_{2})/2$. $\kappa$ is the coupling coefficient between the two lasers, $\tau$ is the time-delay in the coupling that depends on the physical separation between the two lasers, $e^{-i{\Theta}\tau}$ is the phase accumulation due to light propagating from one laser to the other, and $J_{1,2}$ are the injection currents above threshold for each laser. If these four coupled equations can be reduced to two by eliminating the carrier inversion equations as discussed for the single laser with feedback, one is left with two coupled, time-delayed rate equations for the intracavity fields $\mathcal{E}(t)=[E_1(t),E_2(t)]^T$, 
\begin{equation}
\label{eq:2lam}
\frac{d}{dt}{\mathcal E}(t)=M(\frac{d}{dt})\mathcal{E}(t)
\end{equation}
where the $2\times2$ non-Hermitian, time-delay operator $M$ is given by 
\begin{equation}
\label{eq:3lam}
M=
\left[\begin{array}{cc} 
(1+i\alpha)\zeta N_{10}+i\Delta\omega & \kappa e^{-i\Theta\tau} e^{-\tau\partial t}\\ 
\kappa e^{-i\Theta\tau} e^{-\tau\partial t} & (1+i\alpha)\zeta N_{20}-i\Delta\omega,
\end{array}\right]
\end{equation}
and $N_{10}$ and $N_{20}$ are the steady-state carrier inversions. Equations (\ref{eq:2lam}) and (\ref{eq:3lam}) are also amenable to analytic solution via the Lambert W equation formalism.  An experimental study of the bidirectionally coupled lasers, and a detailed comparison between the predictions of the Lambert function solution and numerical solutions of the full Lang-Kobayashi equations will be reported elsewhere.

In summary, we have shown that the Lambert $W$-function provides an hitherto unexplored, analytic method for studying the intensity dynamics of a semiconductor laser with time-delayed optical feedback.  The formalism is valid in a regime where the two rate equations given by the Lang-Kobayashi model are reduced to a single, time-delayed rate equation for the intracavity electric field.  The Lambret function can be invoked when the nonlinearities that arise from gain saturation are neglected, which implies that the analytic results are valid at short times after the laser intensity crosses its threshold value, i.e. when the intensity is still in the exponentially amplifying stage.  Furthermore, the analytic technique assumes that the carrier inversion has reached its steady state value while the intensity is till growing.  To overcome the problem of a very narrow observation time window for the laser intensity dynamics, we have sugggested some remedies that could be implemented at the laser fabrication step to modify the material parameters of the laser.  In particular, reducing the differential gain coefficient, and modifying other properties of the laser so that a very weak pump will induce the desired population inversion, will enable a much wider time window between the time at which the inversion reaches a steady state and the time at which the laser intensity saturates.  Once a wider time window is attained, the predictions of the Lambert function results can be tested.  

%---------------------------------------------------------------------------------------%

%---------------------------------------------------------------------------------------%
%---------------------------------------------------------------------------------------%

\end{document}